\documentclass[twocolumn,preprintnumbers,prl,nofootinbib,floatfix]{revtex4}

\usepackage{graphicx}

\usepackage[hypertex]{hyperref}
\newcommand{\beq}{\begin{equation}}
\newcommand{\eeq}{\end{equation}}

\begin{document}

\pagestyle{plain}

\preprint{VT-IPNAS-10-11}

\title{LSND reloaded}

\author{Sanjib K. Agarwalla}
\email{sanjib@vt.edu}

\author{Patrick Huber}
\email{pahuber@vt.edu}

\affiliation{Department of Physics, Virginia Tech, Blacksburg, VA
  24061, USA}


\begin{abstract}
  In view of the recent result from the anti-neutrino run of
  MiniBooNE, we suggest to repeat the original Liquid Scintillator
  Neutrino Detector (LSND) experiment using Super-Kamiokande, doped
  with Gadolinium, as detector. Due to the more than 100 times larger
  detector mass offered by Super-Kamiokande, the neutrino source
  requires a proton beam power of less than $300\,\mathrm{kW}$ at a
  proton energy around a 1 GeV.  A one year run of this setup can
  corroborate or refute both the LSND and MiniBooNE claims at more
  than $5\,\sigma$ confidence level. If a signal is observed, the
  large size of Super-Kamiokande combined with its good ability to
  determine the position of an anti-neutrino event allows to establish
  the characteristic $L/E$-dependence of oscillation.
\end{abstract}
\maketitle

The LSND
experiment~\cite{Athanassopoulos:1995iw,Athanassopoulos:1997er,Aguilar:2001ty}
has reported a $3.8\,\sigma$ excess of $\bar\nu_e$ events in a beam of
$\bar\nu_\mu$ and recently the MiniBooNE experiment has reported a
$2.8\,\sigma$ excess~\cite{AguilarArevalo:2010wv} of $\bar\nu_e$
events in a beam of $\bar\nu_\mu$ , albeit at a much higher neutrino
energy and at a much longer distance. If one interprets these results
with neutrino oscillation the relevant parameter is the ratio of the
distance $L$ to the neutrino energy $E$, the so called $L/E$. The
$L/E$ ratio is indeed very similar between LSND and MiniBooNE. The
oscillation interpretation of LSND and MiniBooNE points to a mass
squared difference of the order $0.1-10\,\mathrm{eV}^2$ and hence
requires a sterile neutrino. While a sterile neutrino is theoretical
well motivated, the actual oscillation parameters required by the LSND
and MiniBooNE results are in considerable tension with the
non-observation of oscillation in a number of short baseline
disappearance experiments, most notably CDHS~\cite{Dydak:1983zq} and
Bugey~\cite{Declais:1994su}. Also atmospheric and solar neutrino data
show no sign of a sterile neutrino in the required parameter range.
For a recent summary of the status of the oscillation interpretation
of LSND, and by association also of the recent MiniBooNE result, in
the context of all neutrino data, see the review~\cite{Maltoni:2007zf}.

MiniBooNE was designed to be the final test of the oscillation
interpretation of LSND; since MiniBooNE operates at a very different
energy and baseline, only the ratio $L/E$ is similar to LSND and hence
non-oscillation explanations of LSND can not be tested effectively.
So far, MiniBooNE has provided us with the following results:
\begin{enumerate}
\item No oscillation in the neutrino mode for energies above
  $475\,\mathrm{MeV}$~\cite{AguilarArevalo:2007it}
\item An unexplained $3\,\sigma$ excess of $\nu_e$ events in the
  neutrino mode below $475\,\mathrm{MeV}$~\cite{AguilarArevalo:2008rc}
\item A $2.8\,\sigma$ excess of $\bar\nu_e$ events in the
  anti-neutrino mode above $475\,\mathrm{MeV}$, which is consistent
  with LSND~\cite{AguilarArevalo:2010wv}.
\end{enumerate}
In summary, MiniBooNE is not conclusive with respect to the LSND
result and it seems unlikely that a simple increase in statistics
would resolve the issue.  Therefore, the question arises how to
address this problem. One possibility is to repeat an LSND-like
experiment with a pulsed neutrino source~\cite{Garvey:2005pn}, which
requires to build a new liquid scintillator detector.  In
reference~\cite{Grieb:2006mp} it has been proposed to study the
\emph{spatial} dependence of the $\nu_e$ disappearance probability
inside one detector using a radioactive source and in
reference~\cite{Agarwalla:2009em} the same idea was pursued using a
beta-beam anti-neutrino source.

\begin{figure}[t]
\includegraphics[width=\columnwidth]{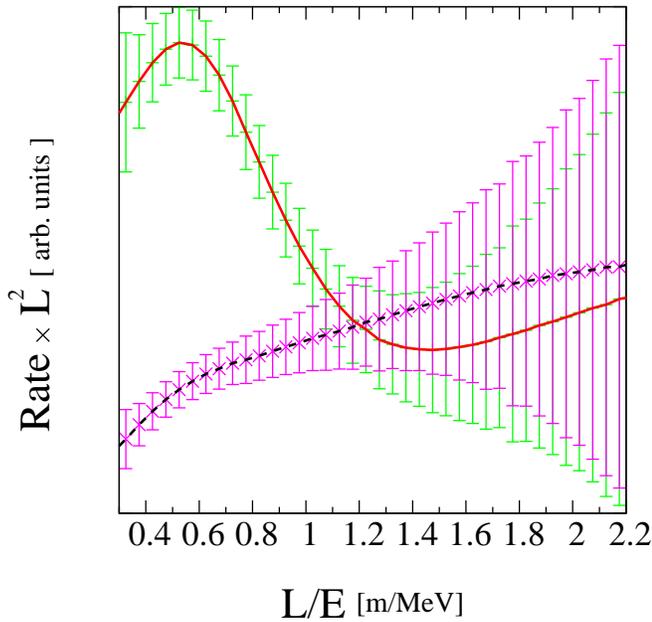}
\caption{\label{fig:baseline} The signal event rate after one year weighted with
  $L^2$ as function of the reconstructed baseline divided by
  reconstructed neutrino energy $L/E$, shown as solid line. The dashed
  line shows the background weighted with $L^2$. The error bars show
  the statistical errors only.  The oscillation signal is computed for
  $\sin^22\theta=10^{-3}$ and $\Delta m^2=2\,\mathrm{eV}^2$.}
\end{figure}

In this letter we suggest to perform a modern version of LSND, {\it
  i.e.} use $\bar\nu_\mu$ from a stopped pion source and inverse beta
decay to detect the appearance of $\bar\nu_e$. The main difference
with respect to the original LSND experiment is that we suggest to use
Super-Kamiokande doped with Gadolinium as
detector~\cite{Beacom:2003nk,Watanabe:2008ru,Dazeley:2008xk} instead
of a liquid scintillator detector. Super-Kamiokande has a fiducial
mass of $22.5\,\mathrm{kt}$ compared to around $120\,\mathrm{t}$ in
LSND. Gadolinium doping allows to efficiently detect the capture of
the neutron which is produced in inverse beta decay. We take 67\% as
detection efficiency which has been obtained from direct tests inside
the Super-Kamiokande detector~\cite{Watanabe:2008ru,Dazeley:2008xk}.
Furthermore, we use an energy resolution as given in
reference~\cite{SK:2008zn} and an energy threshold of
$20\,\mathrm{MeV}$.  Using a detector like Super-Kamiokande has
several advantages. First, the large fiducial mass allows to use a
relatively low power proton source. If we take the same proton source
parameters as in reference~\cite{Conrad:2009mh,Alonso:2010fs} it turns
out that $4\times10^{21}$ neutrino per year are sufficient, which
translates into a proton beam power of only $300\,\mathrm{kW}$. The
contamination with $\bar\nu_e$ from $\pi^-$ decays is very small and
we take a value of
$4\times10^{-4}$~\cite{Conrad:2009mh,Alonso:2010fs}. The neutrino
source will be located on the axis of the cylinder which describes the
fiducial volume and will be $20\,\mathrm{m}$ away from the first
cylinder surface. The resulting signal event rates for one year of
operation are shown in table~\ref{tab:events} and the background event
rate due to beam contamination is $765$.
\begin{table}[h]
\begin{tabular}{|c|cccc|}
\hline
$\Delta m^2$ $[\mathrm{eV}^2]$&0.1&1&10&100\\
\hline
signal&29&1605&1232&1314\\
\hline
\end{tabular}
\caption{\label{tab:events} Number of signal events after one year for 
  $\sin^22\theta=10^{-3}$ including efficiency and energy resolution. }
\end{table}
Secondly, the large rock overburden of approximately
$2,700\,\mathrm{mwe}$, compared to $120\,\mathrm{mwe}$ in LSND,
reduces cosmic ray induced backgrounds to negligible
levels~\cite{Conrad:2009mh,Alonso:2010fs}. Also, atmospheric neutrino
backgrounds are small compared to the beam induced backgrounds.
Thirdly, the large dimensions of the fiducial volume, a cylinder of
$14\,\mathrm{m}$ radius with a height of $36\,\mathrm{m}$ allow to
observe the characteristic baseline dependence of oscillation with
great accuracy.  The size of the copper beam stop used in LSND was
about $50\,\mathrm{cm}$~\cite{Athanassopoulos:1996ds} and the position
resolution for electrons (or positrons) in Super-Kamiokande at
energies above $10\,\mathrm{MeV}$ has been measured to be better than
$75\,\mathrm{cm}$~\cite{Nakahata:1998pz}. Adding these two sources of
baseline uncertainty in quadrature we obtain about $0.9\,\mathrm{m}$.
In our analysis we account for this uncertainty by using a baseline
resolution width\footnote{Obviously, in an actual experiment one would
  include the actual vertex resolution function and distribution of
  pion decays.} of $1\,\mathrm{m}$. We also checked that our results
hardly change for a baseline resolution of $2\,\mathrm{m}$. In order
to be able to perform an $L/E$ analysis we account for an energy
resolution~\cite{SK:2008zn}. Thus, with a source detector distance of
$20\,\mathrm{m}$ and an energy range from $20-52,\mathrm{MeV}$ the
oscillation pattern can be observed for an $L/E$ range of
$0.4-2.8\,\mathrm{m}\,\mathrm{MeV}^{-1}$. This is illustrated in
figure~\ref{fig:baseline}, where we show the signal and background
rates weighted with $L^2$ as a function of $L/E$.  The signal is shown
in red and the background in black. The reason, that we do not see a
simple sine square wave is that the exposure in $L/E$ is non-uniform,
even after rescaling with $L^2$. The oscillation signal is computed
using the usual 2 flavor expression with $\sin^22\theta=10^{-3}$ and
$\Delta m^2=2\,\mathrm{eV}^2$.
Obviously, the $L/E$ dependence is a powerful handle to reject the
background and therefore our results are quite insensitive to
systematic errors. The ability to study the $L/E$ dependence in
detail is crucial if a signal is observed, since it will allow to
establish or refute oscillation as the underlying physical mechanism.
Note, that the this kind of experiment is possible at any large Water
Cerenkov detector and a very similar configuration, although for a
different purpose, has been studied for the detector of the Long
Baseline Neutrino Experiment (LBNE) in
reference~\cite{Agarwalla:2010ty}.

\begin{figure}[t]
\includegraphics[width=\columnwidth]{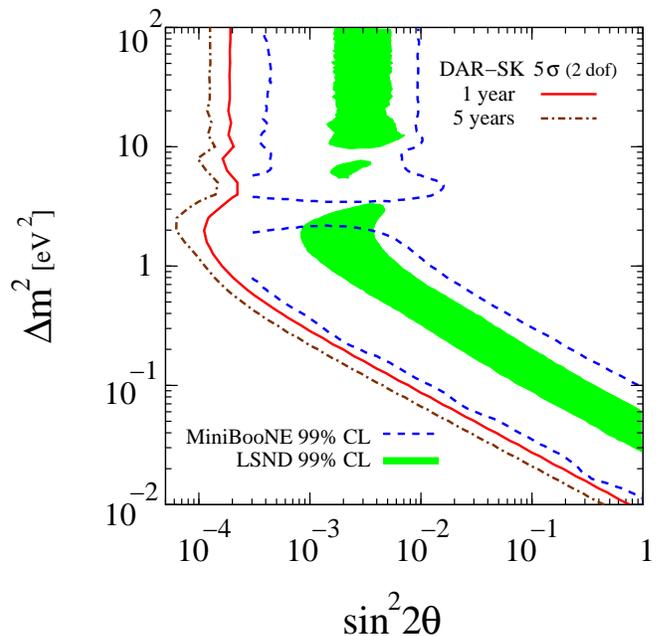}
\caption{\label{fig:sens} Sensitivity limit to sterile neutrino
  oscillation at the $5\,\sigma$ confidence level, shown as red, solid
  line, labeled DAR-SK. This limit corresponds to a one year run. The
  green/gray shaded region is the LSND allowed region at $99\%$
  confidence level, whereas the dashed line is the MiniBooNE
  anti-neutrino run allowed region at $99\%$ confidence
  level~\cite{AguilarArevalo:2010wv}. }
\end{figure}
For our sensitivity calculation, we take systematic errors of 5\% on
both the signal and the background, which are not correlated between
signal and background. They are included using the pull method as
described in {\it e.g.} reference~\cite{Huber:2002mx}. We bin our data
into 38 equally size $L/E$ bins in the range of
$0.4-2.8\,\mathrm{m}\,\mathrm{MeV}^{-1}$. We perform the usual
$\chi^2$ analysis using a Poissonian likelihood function.  Ideally,
one would perform a 2 dimensional binning in both $L$ and $E$, however
the resulting increase in sensitivity would be small since the energy
spread of the beam is relatively small compared to the variation in
$L$. In figure~\ref{fig:sens} we show sensitivity for the $L/E$ binning
analysis at $5\,\sigma$ confidence level (2 degrees of freedom) as
well as the $99\%$ confidence level allowed regions obtained from LSND
and the MiniBooNE anti-neutrino run~\cite{AguilarArevalo:2010wv}.
Note, that the sensitivity is limited by the magnitude of the beam
background and neither increasing the neutrino luminosity or running
time will yield large improvements.  Therefore, our choice of
$4\times10^{21}$ neutrinos at the source is quite optimal.

The experiment we study in this letter can test the LSND and MiniBooNE
claims for $\bar\nu_\mu\rightarrow\bar\nu_e$ oscillations with more
than $5\,\sigma$ significance within one year of running time. The
effort appears moderate: Super-Kamiokande needs to be doped with
Gadolinium and a $300\,\mathrm{kW}$, low energy proton accelerator has
to be installed close to Super-Kamiokande. This setup can provide a
stringent test of previous results due to its high statistics, low
background and the ability to study the baseline dependence in detail.
The baseline dependence also may provide a clue to the underlying
physics in non-oscillation scenarios, which are favored by global
neutrino data~\cite{Maltoni:2007zf}.  The neutrino production and
detection reactions are the same as in LSND and therefore this
experiment will be able to return the final verdict on LSND
irrespective of the underlying flavor transition mechanism.

\acknowledgments

This work has been in part supported by the U.S. Department of Energy
under award number \protect{DE-SC0003915}.

\bibliographystyle{apsrev} 
\bibliography{references}

\end{document}